\documentclass[12pt,showpacs,preprintnumbers,floatfix]{article}
\usepackage{graphicx} 
\usepackage{dcolumn} 
\usepackage{bm} 
\usepackage{amsfonts}
\usepackage{amsthm}
\usepackage{amsmath}
\usepackage{amssymb}
\usepackage{epsfig}
\usepackage{color}
\usepackage{textcomp}
\usepackage{hyperref}
\usepackage{titlesec}
\usepackage{slashed}
\usepackage{caption}
\usepackage{subcaption}
\usepackage{siunitx}
\usepackage{booktabs}
\usepackage{multirow}
\usepackage{cite}
\usepackage{placeins}
\usepackage{comment}
\usepackage{braket}

\def\be{\begin{equation}}
\def\ee{\end{equation}}
\def\ba{\begin{eqnarray}}
\def\ea{\end{eqnarray}}

\def\bdm{\begin{displaymath}}
\def\edm{\end{displaymath}}

\def\bq{\begin{quote}}
\def\eq{\end{quote}}

 at 10truept

\newcommand{\bea}{\begin{eqnarray}}
\newcommand{\eea}{\end{eqnarray}}

\newcommand{\bi}{\begin{itemize}}
\newcommand{\ei}{\end{itemize}}

\newcommand{\beq}{\begin{equation}}
\newcommand{\eeq}{\end{equation}}
\newcommand{\beqa}{\begin{eqnarray}}
\newcommand{\eeqa}{\end{eqnarray}}

\def\ltap{\ \raise.3ex\hbox{$<$\kern-.75em\lower1ex\hbox{$\sim$}}\ }
\def\gtap{\ \raise.3ex\hbox{$>$\kern-.75em\lower1ex\hbox{$\sim$}}\ }
\def\gl{\ \raise.5ex\hbox{$>$}\kern-.8em\lower.5ex\hbox{$<$}\ }
\def\roughly#1{\raise.3ex\hbox{$#1$\kern-.75em\lower1ex\hbox{$\sim$}}}

\begin{document}

\vspace*{2cm}

\begin{center}
{\Large \bf Gravitational Absorption Lines}\\

\vspace*{1.0cm} 
{Andrea Palessandro$^{a}$\footnote{\tt palessandro@cp3.sdu.dk}, 
Martin S. Sloth$^{a}$\footnote{\tt sloth@cp3.sdu.dk}}\\
\vspace{.5cm} {\em $^a$CP$^3$-Origins, Center for Cosmology and Particle Physics Phenomenology \\ University of Southern Denmark, Campusvej 55, 5230 Odense M, Denmark}\\

\end{center}
\begin{abstract}
\noindent 
We consider the gravitational analogue of Lyman-alpha absorption lines in astronomical spectroscopy. If Einstein gravity with minimally coupled matter is valid up to the Planck scale, quantum bound states absorb gravitons of a specific frequency with Planckian cross section, $\sigma_{\text{abs}} \approx l_p^2$. Consequently, one can show that gravitational absorption by bound states is inefficient in ordinary gravity.  If observed, gravitational absorption lines would therefore constitute a powerful smoking gun of new exotic astrophysical bound states (near extremal bound states) or new gravitational physics, as well as give direct evidence of the quantized nature of the gravitational field. We provide, as an example of new gravitational physics near the Planck scale, a non-minimal coupling of the matter fields which breaks the equivalence principle on-shell. We lay out a model in which absorption lines in the primordial gravitational wave spectrum are produced as a consequence of this coupling. 
\end{abstract}

\section{Introduction}

When gravitational waves travel through a medium, they are generally absorbed and re-emitted by the intervening matter. The absorption of gravitational waves in a cosmological setting was first studied by Hawking \cite{Hawking:1966qi}, who calculated the absorption rate of gravitational radiation by viscous matter. Recent studies considered gravitational wave propagation through collision-less matter \cite{Baym:2017xvh,Flauger:2017ged}, as well as quantum mechanical absorption of low frequency gravitational radiation by inverse bremsstrahlung \cite{Flauger:2019cam}. The absorption processes so far considered in the literature all involve the interaction between a graviton and a \textit{scattering state}, that is a quantum state of matter with a continuous energy spectrum. Since the energy of the scattering state can vary continuously, the absorption happens in broad frequency bands.  

In this work we will instead consider the interaction between a graviton and a \textit{quantum bound state} with discrete energy levels. In order to be absorbed, the frequency of the graviton has to match the energy difference between any two quantum states, therefore the absorption will take place in a narrow frequency range, and, if the conditions are right, produce gravitational absorption lines, analogous to their electromagnetic counterparts. We stress that by quantum bound states we do not mean just atoms, which are a particular class of bound states with potential $V(r) \sim r^{-1}$, but all quantum states with a localized wavefunction and a discrete energy spectrum.

Strikingly, all bound states absorb gravitons with the same probability, independently of their internal structure, such as their mass or coupling. This includes purely gravitational atoms \cite{Nielsen:2019izz}, which consist of near Planckian particles bound together by gravity. The remarkable fact that all dimensions other than the Planck length drop out of the gravitational absorption cross section was first shown rigorously in \cite{Boughn:2006st,Rothman:2006fp} in the context of simple atoms. In appendix \ref{gravabs} we show that the result holds for all types of quantum bound states. The absorption cross section is universal, but tiny, of the order of the planck area. As a consequence, the absorption rate for gravitons travelling through the interstellar medium is minuscule, far too small to leave any detectable imprint on gravitational waves of astrophysical origin today. 

The absorption rate is greatly enhanced if the absorbing material is ultradense, like for example in compact stars or in the early universe. Nonetheless, we show that gravitational absorption will not take place even in these extreme cases: a compact star so dense to be on the verge of collapsing to a black hole, as well as a uniform gas of bound states dominating the energy density of the universe at the highest temperatures after inflation, are still too dilute to efficiently absorb gravitons. In fact, one can show that gravitational absorption is always inefficient under very general assumptions, namely 4D Einstein gravity with minimally coupled matter. By inefficient we mean that the optical depth for a graviton traversing any medium is always less than one. The conclusion is very robust, and is independent of the precise expansion history, the composition of the absorbing material, or even the structure of the bound states. As a caveat, we show that gravitational absorption can be marginally efficient (the optical depth reaches order one) if the absorbing material is a maximally dense condensate of a bosonic field with discrete energy levels. We call this bound state \textit{extremal}. An example of a near extremal bound state is the black hole atom of \cite{Arvanitaki:2014wva}. Interestingly, hypothetical black holes with a mass gap \cite{Mukhanov:1986me,Dvali:2011nh} are naturally extremal bound states, i.e. they absorb gravitons with maximum efficiency in Einstein's gravity. 

The insurmountable obstacle in observing gravitational absorption lines actually reflects something deep about the nature of gravity in Einstein's theory. In a classic paper \cite{Bohr:1933zz}, Bohr and Rosenfeld showed that it is mathematically inconsistent to have a classical electromagnetic field interacting with quantum mechanical matter. The argument does not carry through in the same way for gravity: the quantization of the matter fields does not necessarily imply the quantization of the gravitational field. A classical gravitational wave can consistently interact with a quantum measuring apparatus. In fact, not only is the quantization of gravity not a logical necessity in this type of thought experiments, it is also believed to be unobservable within Einstein gravity. Freeman Dyson first showed that it is impossible to detect a single graviton with high probability in any realistic experiment, and conjectured a censorship effect that precludes the observation of the quantization of the gravitational field in Einstein's theory\cite{DYSON:2013jra}. Absorption lines are just another way of probing the quantization of gravity, therefore they are excluded by similar arguments. 

Absorption lines become possible if minimally coupled Einstein gravity is modified at high energies. As an example, we consider a non-minimal coupling to gravity which breaks the equivalence principle for the particles in the bound state. By tuning the non-minimal coupling parameter, one can greatly enhance the absorption cross section. Absorption lines in non-primordial spectra due to non-minimally coupled fields require light particles with strong interactions, which are excluded by LHC constraints. However, there are in principle no obstructions to absorption lines in primordial gravitational spectra. In particular, we describe a scenario in which massive particles interacting non-minimally with gravity decouple from the hot plasma shortly after inflation and quickly become non-relativistic, eventually recombining in much the same way as ordinary hydrogen atoms. The newly formed atoms absorb primordial gravitons of a specific frequency, leading to a series of absorption lines in the primordial gravitational wave spectrum. 

Dyson's work was motivated by the hope that the failures in reconciling General Relativity with Quantum Mechanics were really due to the fact that the gravitational field is a purely classical entity. In the dichotomous world he envisioned, the geometric theory of gravity would peacefully coexist with the quantum realm, and the obstruction in observing individual gravitons would be attributable to their non existence. Although we believe it will be very difficult to have a consistent effective quantum field theory description of Nature below the Planck scale without quantizing gravity and thus introducing a graviton\footnote{We expect to return to this point in future work.}, gravitational absorption lines would provide a way to discriminate between classical and quantum gravity at the observational level. In fact, gravitational absorption lines would probe quantum gravity in two distinct ways. First of all, since General Relativity forbids them, they would explore and constrain exotic physics close to the Planck scale. Secondly, and perhaps more importantly, they would confirm that the gravitational field is quantized at low energies, effectively proving the existence of gravitons.

\section{A no-go argument}
We now show that under very general assumptions about the theory of gravity, gravitational absorption is always inefficient, meaning that it is impossible for a graviton to be absorbed with high probability by any kind of bound state. This is the main conclusion of this section. The assumptions are the following:

\begin{itemize}
    
  \item Einstein gravity is valid up to the Planck scale. 
  
  \item All fields are minimally coupled to gravity. 

  \item The absorbing material is made up of bound states with a discrete energy spectrum. 
  
\end{itemize}

We define ``bound state" as any type of localized quantum state with discrete energy levels. Atoms are a particular subclass of bound states with an inverse square law potential, but the conclusions of this section apply more generally to any type of confining potential. The absorption cross section for the transition between the 1s and 3d state of a hydrogen-like atom \footnote{When the atom absorbs a graviton, it transitions from the $n=1$ ground state to the first excited state with $l=2$, namely $n=3$.} was first computed in \cite{Boughn:2006st},
\begin{equation}\label{abssigma}
\sigma_{abs}=\frac{3^4 \pi^2}{5\times 2^9} \, l_p^2 \approx 0.3 \, l_p^2,
\end{equation}
where $l_p$ is the 4d planck length. Strikingly, all dependence on the mass and coupling drops out and the final cross section is comparable to the planck area within a numerical factor of order one, a fact emphasised also in \cite{Rothman:2006fp,DYSON:2013jra}. Detailed calculations leading to (\ref{abssigma}), as well as some heuristic arguments for why this is true, can be found in appendix \ref{gravabs}. In the appendix we also extend the result to multi-particle atoms, nuclei, and finally to all quantum states with a confining potential, reaching the conclusion that any (non-degenerate) bound state will absorb with a planck area cross section. Degenerate bound states with large occupation numbers have a larger cross section, but are proportionally heavier, therefore they do not constitute an exception to the no-go, as we explain in section \ref{caveat}. 

Technically, we will say that gravitational absorption is efficient whenever the optical depth for a graviton going through a material is strictly larger than one. If the optical depth is equal to one we will say that the material is marginally efficient at absorbing gravitons. We will first prove the inefficiency of gravitational absorption in the framework of the standard cosmological scenario. Then, we will formulate our conclusion in the larger context of Dyson's conjecture and prove it more generally. 

\subsection{Heuristic no-go in standard cosmology}

Here we assume that Friedmann equations govern the evolution of the universe at all times, which is driven by ideal fluids with equation of state $p=\omega \rho$ and $\omega \leq1$, so that the speed of sound $c_s=\sqrt{\partial p/ \partial \rho}=\sqrt{\omega}$ is subluminal. The energy density scales as $a^{-3(1+\omega)}$ and it can decrease at most like $a^{-6}$ for a stiff fluid with $\omega=1$. While these extra assumptions are needed for the present discussion, we will see in the next section that our conclusion is actually stronger and only depends on the nature of gravity itself. 

We can consider two types of gravitational waves, primordial and non-primordial, depending on whether they are created close to inflation or much later. Non-primordial gravitational wave sources are typically of astrophysical origin, such as binary systems, supernovae and spinning neutron stars. Different processes in the early Universe may have generated a primordial gravitational wave background, such as, among others, quantum perturbations during inflation \cite{Maggiore:2000gv,Pagano:2015hma}. 

The no-go argument comes in two parts, depending on the character of the gravitational wave signal. We start with the non-primordial component. Assuming that the absorption is due to gas clouds in the interstellar medium, a necessary condition to have absorption lines in non-primordial gravitational waves is that the optical depth for a graviton travelling through the galaxy is larger than one,
\begin{equation}\label{nonprimordial}
n_{B} \sigma_{abs} R_G > 1 ,
\end{equation}
where $R_G \sim 10 \, \text{kpc}$ is the radius of our galaxy. This gives a lower bound on the number density of bound states today: 
\begin{equation}\label{nonprim}
n_{B} > (\sigma_{abs} R_G)^{-1} \sim 10^{-54} m_p^3,
\end{equation}
at least forty orders of magnitude larger than the density of hydrogen atoms in the interstellar medium. Even if we assume that dark matter is mostly comprised of bound states with mass $m_B$ and a number density given by (\ref{nonprim}), for such a dense gas not to overclose the universe, the mass should be incredibly small, $m_B \lesssim 10^{-68} \, m_p$. The Compton wavelength of a single particle $\lambda_B = m_B^{-1}$ would then be 10 million times larger than the Hubble radius today, and the particle description would break down.  

What about absorption by super-dense compact objects (e.g. neutron stars)? Calling the mass and radius of the compact star $M_S$ and $R_S$, the condition for absorption is
\beq\label{compact}
n_{B} > \frac{m_p^2}{R_S}.
\eeq
Bound states of mass $m_B$ need to be confined inside the star, so their size should be much smaller than the radius of the star. Therefore, at the very least
\beq
m_B > R_S^{-1}. 
\eeq
The energy density of bound states is $\rho_B =m_B n_B > m_p^2/R_S^2$, so the total mass of the star is at least $\rho_B R_S^3 > m_p^2 R_S$, i.e. $M_S > m_p^2 R_S$. However, the mass of the star has to be below the corresponding mass for a black hole of that size, namely $M_S < m_p^2 R_S$, a contradiction. No compact star, no matter its composition, will absorb gravitons with high probability. 

Analogously, the necessary condition for gravitons from the primordial spectrum to be absorbed in the early universe is that the absorption rate is larger than the Hubble rate:
\begin{equation}\label{primordial}
n_B \sigma_{abs} H^{-1}> 1.
\end{equation}
The difference between this and the previous case is that the size of the system $H^{-1}$ is now changing with time. The number density of bound states scales like $n_B \propto a^{-3}$, while the Hubble rate scales like $H \propto a^{-3(1+\omega)/2}$, depending on which fluid dominates the early evolution of the universe. So, $n_B$ decreases faster than $H$ for all values of $\omega < 1$. In the special case of a stiff fluid, the ratio between the number density and Hubble rate remains constant. 

This means that absorption will be most efficient at the earliest time after formation of these bound states. If condition (\ref{primordial}) is not satisfied immediately after bound state formation, it will never be satisfied. Assuming that the constituent particles were in thermal equilibrium with the SM plasma before decoupling and subsequent recombination at temperature $T$, the bound state number density is bounded by the equilibrium number density of the particles in the plasma and therefore has to satisfy 
\beq
n_B < \frac{2 \zeta(3)}{\pi^2} T^3.
\eeq
The Hubble rate in a radiation dominated universe is given by
\beq
H^2 = \frac{8 \pi^3}{45 \, m_p^2} g_{SM} T^4, 
\eeq
where $g_{SM}$ is the number of relativistic degrees of freedom in the visible sector at temperature $T$. Condition (\ref{primordial}) is then impossible to realise for $T< m_p$.

Note that equations (\ref{nonprimordial}), (\ref{compact}) and (\ref{primordial}), which place a lower limit on the number density of bound states such that gravitational absorption is efficient, only depend on fixed parameters like the typical radius of a galaxy, the size of a compact star, or the Hubble rate. All information about the bound state, such as its mass or coupling, is irrelevant. This is due to the universal nature of the gravitational absorption cross section (\ref{abssigma}): all bound states absorb gravitons with the same probability, regardless of their structure. Efficient gravitational absorption is nevertheless impossible in the examples that we discussed. Is there a deeper reason for this? As it turns out, there is: absorption lines are forbidden on general grounds if the theory of gravity is General Relativity. Detecting absorption lines is really the same as proving the existence of gravitons and it appears that Einstein's theory somehow conspires to hide the quantization of the gravitational field. 

\subsection{Relation to Dyson's conjecture}
Freeman Dyson first pointed out that the quantization of the gravitational field is not a logical consequence of the quantum behaviour of matter. This is in stark contrast with the electromagnetic case, where one can indeed show that a classical electromagnetic wave interacting with quantum matter would lead to inconsistencies \cite{Bohr:1933zz}. Dyson then asked the question whether it is in principle possible to detect the quantization of the gravitational field \cite{DYSON:2013jra}, and found that this is impossible if one uses atoms as detectors, due to the planck area cross section for absorption. In appendix \ref{gravabs} we are able to extend the result, and therefore Dyson's conclusion, to all types of bound states. 

We briefly review Dyson's argument. Independently of the precise nature of the experimental apparatus, in order to detect a single graviton with high probability the size of the detector $R$ should exceed the mean free path of the graviton, or
\beq \label{Dyson}
n_B \, \sigma_{\text{abs}} R \geq 1,
\eeq
where $n_B$ is the number density of detector particles and $\sigma_{\text{abs}} $ the absorption cross section for gravitons. The detector particles in our case are bound states with mass $m_B$. If the detector has mass M, the number density of bound states is $n_B = M/(m_B R^3)$, within numerical factors of order one. We can then write condition (\ref{Dyson}) as 
\beq\label{Dysoncond}
\frac{M}{R} \frac{l_p^2}{m_B R} \geq 1.
\eeq 
Now, $M/R \lesssim m_p^2$ for any object that is not a black hole, therefore condition (\ref{Dysoncond}) requires $m_B R \lesssim 1$. However, at the very least the Compton wavelength of a single particle should be smaller than the size of the detector, $m_B^{-1} \lesssim R$. The two conditions are incompatible. This constitutes a strong indication that is in principle impossible to detect a single graviton with high probability using bound states as detectors, and consequently that any atomic (or nuclear) gas will never be dense enough to produce absorption lines. 

Note that condition (\ref{Dyson}) is identical to conditions (\ref{nonprimordial}) and (\ref{primordial}), with $R$ being the typical radius of a galaxy $R_G$ and the Hubble radius $H^{-1}$, respectively. What this means is that we can effectively treat a galaxy or the entire universe as graviton detectors and whether these systems constitute good graviton detectors basically depends on whether they are efficient at absorbing gravitons. Dyson showed that condition (\ref{Dyson}) is never satisfied in any system obeying Einstein gravity, therefore, if true, it also forbids gravitational absorption lines. 

Explicitly, if we consider a detector as big as the observable universe, containing a dense soup of bound states with the maximum possible number density $n_B^{\text{max}}=(3/8 \pi) H^2 m_p^2/m_X$, the inequality (\ref{Dysoncond}) reads $H \gtrsim m_B$. Again, the localization of the wave function requires $m_B \gtrsim H$, in contradiction with the previous inequality. We conclude that the condition can never be satisfied, even if we fill the observable universe with the highest possible number density of bound states, regardless of how they are created, their morphology, or their cosmological evolution. The only assumption our conclusion rests upon is that Einstein's theory of gravity, with minimally coupled matter, is valid.  

\section{A caveat: extremal bound states}\label{caveat}

One could in principle relax condition (\ref{Dyson}) by requiring that only an order one fraction of gravitons be absorbed by the detector.  For example, if the graviton mean free path was twice the radius of the detector, we should still expect a sizeable fraction of gravitons to be absorbed.  That could be enough to leave a detectable imprint on gravitational wave signals under extreme conditions. 

Suppose we have a quantum bound state of mass $M$ and size $R$ made up of particles with mass $m_B$. The constituent particles are the ones that undergo quantum transition when a graviton is absorbed. Gravitational absorption is most efficient when the density is maximal. Assuming that the system is on the verge of gravitational collapse, $M/R\sim m_p^2$, condition (\ref{Dyson}) is satisfied when $m_B R \lesssim 1$. On the other hand, we need $m_B R \gtrsim 1$, otherwise the system would be smaller than the Compton wavelength of its constituents. The two conditions are both marginally satisfied whenever
\beq\label{extremal}
m_B R \sim 1,
\eeq
namely when the Compton wavelength of a singe particle is as large as the whole bound state. A maximally dense quantum bound state made up of particles that satisfy (\ref{extremal}) is marginally efficient at absorbing gravitons. We call such a physical system an \textit{extremal bound state}. We stress that extremal bound states are not only maximally dense, but also, in a sense, maximally delocalized, as the constituent particles are as large as the system itself. In particular, this means that the constituent particles have to be bosons, otherwise Pauli's exclusion principle would prevent more than one particle from occupying the same quantum state.  

An example of an (almost) extremal bound state is the ``gravitational atom" of \cite{Arvanitaki:2014wva}. Ultralight bosons can induce superradiant instabilities in spinning black holes \cite{Penrose:1969pc,Zouros:1979iw}, tapping their rotational energy to trigger the growth of a bosonic condensate that binds to the black hole ``nucleus" in a macroscopic quantum bound state (more conventional quantum bound states of heavy fundamental particles held together by gravity were considered in \cite{Nielsen:2019izz}, and are also called gravitational atoms). 

The gravitational coupling constant $\alpha_G$ of the black hole atom is given by the ratio between the gravitational radius of the black hole and the Compton wavelength of the bosonic field:
\beq
 \alpha_G = \frac{R_{BH}}{2 \lambda_C} = G M_{BH} m_B,
\eeq
where $M_{BH}$ is the mass of the black hole, and $R_{BH}$ its radius. The bosons form a cloud at a distance 
\beq
r_C \sim \frac{R_{BH}}{\alpha_G^2}
\eeq
from the black hole. Unless $\alpha_G$ is very close to one, $r_C \gg R_{BH}$, and the size of the bound state is given by the extent of the boson cloud $r_C$. The superradiant condition requires $\alpha_G < 1$, therefore $r_C>\lambda_C>R_{BH}$, namely the Compton wavelength of the bosons is smaller than the size of the bound state, which is therefore less than extremal. The black hole atom only becomes extremal in the limit $\alpha_G \rightarrow 1$, in which the boson cloud collapses on the horizon, $r_C \sim R_{BH}$, and the Compton wavelength of the bosons becomes equal to the size of the bound state. One can even show that the superradiance rate for spinning black holes is maximized when the Compton wavelength of the massive bosonic particles is comparable to the black hole size. 

There is another way to see that the black hole atom is at best extremal. Following the triggering of the instability, the number of bosons occupying the ground state of the atom grows exponentially, extracting energy and angular momentum from the black hole. The growth stops when a fraction $\Delta L \sim \mathcal{O}(0.1)$ of angular momentum has been extracted from the black hole, leading to occupation numbers of the order 
\begin{equation}
N \simeq G M_{BH}^2 \Delta L \sim \left(\frac {M_{BH}}{m_p}\right)^2 \Delta L.
\end{equation} 
The bosons (with mass $m_B$) have now a Compton wavelength that is comparable to the size of the black hole, therefore $m_B \approx (G M_{BH})^{-1}$. Since the particles making up the atom are bosons, the cross section (\ref{abssigma}) is enhanced by the Bose-Einstein factor,
\begin{equation}\label{absBH}
\sigma_{\text{abs}} \approx 0.3 (1+N) \, l_p^2,
\end{equation}
where $N$ is the occupation number of the ground state. If $N \gg 1$, the absorption cross section is much larger than the planck area. Condition (\ref{Dyson}) now reads 
\beq
n_{BH} \sigma_{\text{abs}} R_{BH} \geq 1,
\eeq
where $n_{BH} \sim R_{BH}^{-3}$ is the number density of the black hole bound state (in this picture, there is exactly \textit{one} bound state, the black hole), $R_{BH}=2GM_{BH}$ its Schwarzschild radius, and $\sigma_{\text{abs}}$ is given by (\ref{absBH}), where one can neglect the first term since $N \gg 1$. The condition then just becomes $\Delta L \geq 1$, which is only satisfied in the extremal case $\Delta L =1$, in which all of the angular momentum of the spinning black hole is extracted. In particular, note that Dyson's conjecture still holds, since efficient absorption would entail $\Delta L >1$, an impossibility. 

Black hole superradiance is not the only process in which the absorption cross section can be enhanced via the Bose factor. Any Bose-Einstein like condensate with large occupation number will have an enhanced cross section. If the ground state of each bound state is populated by $N$ particles of mass $m_B$, the cross section is enhanced by  $\sigma_{\text{abs}} = N l_p^2$. At the same time, however, the total mass of a single bound state proportionally increases by the same factor, therefore the number density of bound states in a detector of total mass $M$ decreases to $n_B=M/(N m_B R^3)$, and condition (\ref{Dysoncond}) is left unchanged. Intuitively, the reason why Bose enhancement does not help with the no-go is that one cannot take very large values of $N$, such that absorption is efficient, without collapsing the whole system into a black hole. The best one can do is to saturate condition (\ref{Dyson}) by having a bosonic condensate with discrete energy levels that is both maximally dense and maximally delocalized, in the sense we explained previously.

One might also speculate about hypothetical physical systems that naturally saturate condition (\ref{Dyson}), and are therefore perfectly extremal: black holes with a mass gap. It has been argued that in any sensible quantum field theory the mass of a black hole must be quantized \cite{Mukhanov:1986me,Dvali:2011nh}. If this is true, black holes cannot emit or absorb arbitrarily soft quanta, and they effectively act as extremal bound states. A very simple way of seeing this is as follows. Treating the black hole as a single macroscopic bound state of size $R_{BH}$, its number density is simply $n_{BH} \sim 1/R_{BH}^3$. A black hole absorbs everything that comes into contact with its event horizon, therefore its absorption cross section is as large as the area of the horizon
\beq\label{BHabs}
\sigma_{\text{abs}}^{BH} = 4 \pi R_{BH}^2.
\eeq
The optical depth for a graviton going through the black hole is therefore
\beq
n_{BH} \sigma_{\text{abs}}^{BH} R_{BH} \sim 1.
\eeq
Incidentally, one can derive the absorption cross section for a black hole (\ref{BHabs}) also from the previous example of the black hole atom, in the appropriate limit. In that scenario the absorption cross section for each particle is Planckian. In the limit $\alpha_G \rightarrow 1$, the boson cloud collapses into the black hole, and the occupation number becomes $N \rightarrow (M_{BH}/m_p)^2 \sim R_{BH}^2 m_p^2$. The effective absorption cross section of the whole system, which is now a black hole, is then $\sigma_{\text{abs}}^{BH}= N \sigma_{\text{abs}}\sim R_{BH}^2$.

It is interesting to compare this with the quantum portrait view of a black hole as a Bose-Einstein condensate of soft gravitons with large occupation numbers \cite{Dvali:2011aa}. In this picture the black hole is a condensate of gravitons of wavelength $R_{BH}$, and occupation number $N \sim (R_{BH}/l_p)^2$. The absorption cross section for exciting a single graviton is Planckian, but, since there are $N$ gravitons in the lowest level and any of them can be excited, there is an extra factor of $N$, so the resulting cross section is of order\footnote{We thank Gia Dvali for correspondence clarifying this point.} $R_{BH}^2$. 

While a quantized black hole could absorb gravitons semi-efficiently, it is unclear whether this will show in the spectrum as absorption lines. Indeed, it was argued in a series of papers starting from \cite{Dvali:2012en} that the level spacing for an ordinary black hole must be absolutely minuscule. Namely, it must be of order $\lesssim (S_{BH}R_{BH})^{-1}\sim m_p^4/M_{BH}^3$, where $S_{BH}=\pi (R_{BH}/l_p)^2$ is the entropy of the black hole. In this picture, black holes are quantum states with large occupation numbers, and are therefore effectively classical. As a result, the quantum levels are very closely spaced, such that the spectrum appears almost continuous. Nevertheless, the level spacing increases for smaller black holes, so that microscopic black holes could have a sizeable spacing. For example, a black hole with a mass of  $\sim 10^6$ kg would have a level spacing in the Hertz range, and in particular it could only absorb particles with a frequency that is a multiple of the Hz. In the limit of a Planckian black hole, the level spacing would also be Planckian, $\Delta E \sim m_p$.

\section{Evading the no-go}
The no-go argument we presented relies on Einstein's gravity with minimally coupled fields. What happens if we relax one of these assumptions? As it turns out, the no-go argument is remarkably robust, holding even if the gravitational coupling strength is changed, or if a simple non-minimal coupling for $X$ is introduced. Nevertheless, it is possible to evade the conclusions of the argument in certain exotic scenarios, as we will show in this section.

First of all, naively increasing the strength of gravity does not work. If we make the planck length larger, the absorption cross section increases, but the maximum number density of bound states before gravitational collapse ensues proportionally decreases, so that the final absorption rate stays the same. Concretely, the absorption cross section goes like $\sigma_{\text{abs}} \sim 1/m_p^2$, while the maximum number density goes like $n_{B,\text{max}} \sim m_p^2/(m_B R^2)$, where $m_B$ is the atomic mass and $R$ the size of the detector. The two quantities scale in opposite ways with $m_p$, so that the maximum absorption rate $\Gamma_{\text{max}} = \sigma_{\text{abs}} n_{B,\text{max}} $ is independent of the planck mass and increasing or decreasing the gravitational coupling does not have any effect. 

We conclude that any model whose only effect is to change the fundamental scale at which gravity becomes non-perturbative is not going to help. These include, for example, all scenarios with large extra dimensions (LED) \cite{ArkaniHamed:1998rs}. In these models the Planck scale is not fundamental, and its enormous value is simply a consequence of the large size of the extra dimensional space. The ``true" scale of gravity can be much lower, for instance of the order of the electroweak scale for LED models that solve the hierarchy problem. At low energies the extra dimensions are hidden, and gravity is weaker compared to the other forces because the gravitational flux also spreads in the extra dimensions. At high energies, however, the extra dimensions are resolved and the fundamental gravity scale restored to its true value. In a (4+n)-dimensional spacetime with $n$ compactified extra dimension of volume $\mathcal{V}$, the fundamental gravity scale $M_P$ is related to the usual Planck scale via
\beq
m_p^2 = M^{2+n}_P \mathcal{V}. 
\eeq
The current large value of $m_p$ (thus small value of $G$) is simply due to the large volume $\mathcal{V}$ of the extradimensional space. One can then envision a scenario in which $\mathcal{V}$ was much smaller in the early universe, effectively making gravity much stronger at that epoch. This however does not make gravity more efficient at absorbing gravitons, as we saw, since a stronger gravity also makes it that much easier to create black holes. This is a scenario of modified gravity that simply changes the gravity scale at high energies, and as such it cannot work. 

Similarly, introducing a non-minimal coupling for $X$ of the form $\xi R X^2$ does not help. The lowest order vertex connects a single graviton line to two $X$ lines, so it describes $X$ radiating off a single graviton, or, in the time-reversed process, a single graviton being absorbed by $X$. To leading order in $m_p$ the Ricci scalar is $R \sim \Box h/m_p$, where $h_{\mu \nu}$ is the linearized metric. Thus, in the transverse-traceless (TT) gauge ($h^\mu_\mu \equiv h=0$, and $\partial^\mu h_{\mu \nu} = 0$) the term $\xi R X^2$ gives a contribution to the absorption amplitude proportional to $\Box h=0$. Since the amplitude is invariant, the non-minimal contribution vanishes in all gauges. The same is true of all non-minimal couplings involving the Ricci tensor, such as $(\xi/m_p^2) R_{\mu \nu} \partial ^\mu X \partial^\nu X$. The Ricci tensor $R_{\mu \nu}$ only contains terms like $\Box h_{\mu \nu} \sim (p_\alpha p^\alpha) h_{\mu \nu}=0$, that vanish for a graviton on-shell, or terms like $\partial_\mu \partial_\nu h$ and $\partial_\mu \partial^\alpha h_{\alpha \nu}$ that vanish in the TT-gauge. Amplitudes derived from this coupling with an external graviton on-shell are therefore zero in all gauges. 

While the Ricci scalar and the Ricci tensor both vanish in vacuo, the full Riemann tensor does not in general. Consequently, a non-minimal coupling of the form\footnote{Incidentally, this operator was also mentioned in \cite{Solomon:2017nlh} as a higher derivative operator beyond the Horndenski class, whose phenomenological implications should be investigated.}
\begin{equation}\label{nonmin} 
\frac{\xi}{m_p^4} R_{\mu \nu \alpha \beta} \partial^\mu \partial^\alpha X \partial^\nu \partial^\beta X ,
\end{equation}
where $\xi$ is a dimensionless parameter, may give a non-zero contribution to the absorption cross section. We prove that this is the case in appendix \ref{gravabs}.  In a nutshell, the reason is that while $R$ and $R_{\mu \nu}$ only contain terms like $\Box h$, $\Box h_{\mu \nu}$, $\partial_\mu \partial_\nu h$ and $\partial_\mu \partial^\alpha h_{\alpha \nu}$, that all vanish on-shell due to gauge invariance of the gravity action, the Riemann tensor contains non-vanishing terms like $\partial_\mu \partial_\nu h_{\alpha \beta}$. In general, vacuum solutions of Einstein field equations require $R_{\mu \nu}=0$, and as a consequence $R=0$, but the Riemann tensor can be non-zero. 

All non-minimal coupling terms break the equivalence principle, as they introduce an additional coupling between the gravity and matter sectors, but (\ref{nonmin}) is the only one that does not automatically vanish on shell for a graviton absorption process. The central conclusion of \cite{Boughn:2006st,Rothman:2006fp} that the absorption cross section should be planckian rests on two major assumptions: Lorentz invariance and the equivalence principle. The operator (\ref{nonmin}) preserves Lorentz invariance but it explicitly breaks the equivalence principle, so in general one shouldn't expect the resulting cross section to be planckian.

Note that the term $R_{\mu \nu \alpha \beta} \partial^\mu \partial^\nu X \partial^\alpha \partial^\beta X$ would be trivially zero because of the skew symmetry of the Riemann tensor, $R_{\mu \nu \alpha \beta}=-R_{\nu \mu \alpha \beta}=-R_{\mu \nu \beta \alpha}$, whereas the term in (\ref{nonmin}) is not since its index structure does not exhibit any definite symmetry under the exchange $\mu \leftrightarrow \nu$ or $\alpha \leftrightarrow \beta$. In the following, we will assume for simplicity that the bound states are 2-particle atoms in their fundamental energy level.

The absorption cross section, corrected by the new term, is (equation (\ref{sigmaxi}) of the appendix)
\beq\label{sigmaX}
\sigma_{\text{abs}} = \frac{3^4 \pi^2}{5 \times 2^9} G \left[1 + \frac{2^{10}}{3^8 \times 5^2} \xi^2 \left( \frac{m_X}{m_p}\right)^8 \alpha_X^{8}  \right].
\eeq
The highest possible value of $\xi$ compatible with unitarity is $\xi_{\text{max}} \sim m_p^5/(m_X^3 k^2)$. Measuring $\xi$ in units of $\xi_{\text{max}}$, $\tilde{\xi}\equiv \xi/\xi_{\text{max}}$, we get for the cross section
\beq
\sigma_{\text{abs}} = \tilde{\xi}^2 \frac{\alpha_X^4}{m_X^2}. 
\eeq
We can now revise our absorption arguments with the new cross section. 

The condition for absorption (\ref{Dyson}) gives
\beq
\frac{M}{R} \frac{\tilde{\xi}^2 \alpha_X^4}{m_X^3 R} >1,
\eeq
where $M$ and $R$ are the mass and size of the detector. As before, we need $M/R< m_p^2$ to avoid gravitational collapse (the Schwarzchild solution is a vacuum solution and we do not expect it to be affected at the classical level by the non-minimal coupling), and the atoms have to be contained inside the detector, therefore $\alpha_X m_X > R^{-1}$. An additional requirement is that the wavelength of the absorbed graviton is smaller than the size of the detector. Since the frequency of the graviton is of order the binding energy of the atom, this gives the stronger condition $\alpha_X^2 m_X > R^{-1}$. Putting the two together, we obtain
\beq\label{conditionR}
\frac{1}{m_X \alpha_X^2}< R < \tilde{\xi}^2 \alpha_X^4 \frac{m_p^2}{m_X^3},
\eeq
which can be satisfied by a careful choice of parameters. In particular, the interval in (\ref{conditionR}) is non-empty for $m_X<\alpha_X^3 \tilde{\xi} m_p$. 

In a concrete example, imagine the universe to be dominated by non-minimally coupled atoms. The condition for absorption is
\beq
n_B^{\text{max}} \sigma^{\text{max}}_{\text{abs}} H^{-1} >1,
\eeq
where $n_B^{\text{max}}=(3/8 \pi) H^2 m_p^2/m_X$ is the maximum allowed number density of atoms in a universe with Hubble rate $H$. The bound on the Hubble rate is
\beq
H\gtrsim \frac{1}{\tilde{\xi}^2 \alpha_X^4} \frac{m_X^3}{m_p^2},
\eeq
while the bound on the graviton frequency is $\alpha_X^2 m_X > H$. Putting the relevant constraints together we get
\beq\label{Hbound}
\frac{m_X^3}{m_p^2} \frac{1}{\tilde{\xi}^{2} \alpha_X^4}<H<\alpha_X^2 m_X .
\eeq
Again, the interval is non-empty for $m_X<\alpha_X^3 \tilde{\xi} m_p$.

If we want the structure of the atoms to be unaffected by the new coupling, the gauge force has to dominate the interaction between $X$ particles, therefore we need roughly (see the scattering amplitude of (\ref{MSt})) 
\beq\label{alphabound}
\alpha_X \gtrsim \frac{\xi^2}{m_p^{10}} m_X^4 k_B^6 \equiv \tilde{\xi}^2,
\eeq
where $k_B = \alpha_X m_X$ is the Bohr momentum of the particles in the atom. Clearly, if $\xi$ is equal to its maximum value, the right hand side of (\ref{alphabound}) is of order 1 and the condition cannot be satisfied. This is just telling us the obvious fact that if we saturate the unitarity bound non-minimal gravitational interactions will dominate. The parameter $\xi$ has to be large enough to induce gravitational absorption, but small enough to avoid overcoming the gauge forces inside the atom. Equation (\ref{alphabound}) then simply constitutes a further constraint on the model, in addition to (\ref{Hbound}).

\begin{figure}
    \centering
    \includegraphics[width=12cm]{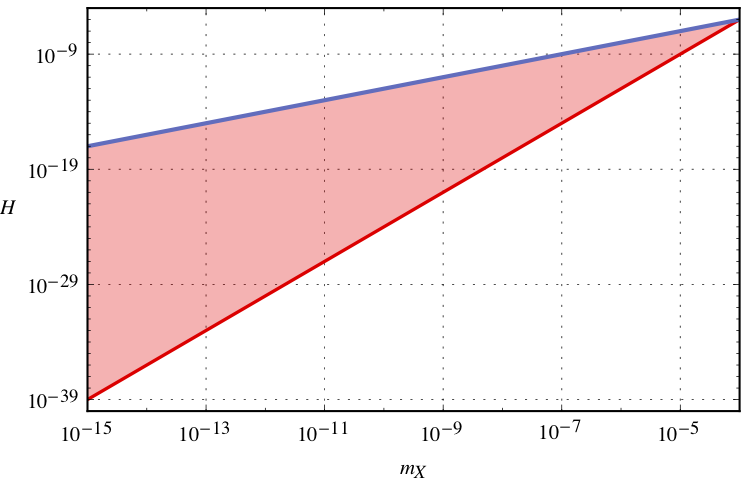}
    \caption{The shaded red region represents the range of values of $H$ that are consistent with gravitational absorption lines if the universe is dominated by non-minimally coupled atoms with mass $m_X$, and $\alpha_X=\tilde{\xi}=0.1$. The red line represents the lower limit $10^6(m_X^3/m_p^2)$, while the blue line represents the upper limit $10^{-2} m_X$. The H range widens as the atomic mass decreases. All quantities are in Planck units.}
    \label{plot1}
\end{figure}

For example, taking $\tilde{\xi} \sim 0.1$, and $\alpha_X \sim 0.1$, (\ref{alphabound}) is automatically satisfied, while (\ref{Hbound}) becomes $10^6 (m_X^3/m_p^2) <H<10^{-2} m_X$. The interval then is non-empty for $m_X \lesssim 10^{-4} m_p$. If the mass saturates the bound, the interval closes around $H \sim 10^{-6} m_p$, which is the current upper limit on the Hubble rate coming from the non observation of tensor modes in the CMB. Efficient gravitational absorption today, on the other hand, is only reached for $m_X \lesssim 10^{-22} m_p$, of the order of the electron mass or smaller. Unfortunately, particles this light, and that interact so strongly with gravity, would have been detected in particle accelerators by now. In fact, the bound on the mass coming from collider searches for $\tilde{\xi}$ close to one is just given by the energy threshold at LHC, $m_X \gtrsim 10 \, \text{TeV}$ (see appendix \ref{gravabs} for details). For this reason, gravitational absorption in this scenario is only viable in the very early universe. Fig.\ref{plot1} shows the allowed range for H as a function of the mass $m_X$, for $\tilde{\xi}=\alpha_X=0.1$, and $m_X > 10^{-15} m_p$. 

So far we just showed that it is possible in principle to tune the non-minimal coupling parameter to extremely high values in order to efficiently absorb gravitons. We will now describe a specific scenario in which gravitational absorption lines are produced as a consequence of this. As we saw, our model is only viable for $m_X \gtrsim 10 \, \text{TeV}$, so for $H \gtrsim 10^{-39} m_p$ (see Fig.\ref{plot1}), in the very early universe. We can then imagine a scenario of the following sort:
\begin{enumerate}
\item Cosmic inflation generically predicts a primordial background of gravitational waves with a flat spectrum. 
\item A massive field $X$ is non-minimally coupled to gravity through the term (\ref{nonmin}). The field is also unstable, and decays to radiation after a typical lifetime that is larger than the (gravitational) absorption time. Excitations of the field are initially in thermal equilibrium with the SM plasma. 
\item We assume that the non-minimal coupling parameter $\xi (\psi)$ depends on the value of some scalar field condensate $\psi$, and is initially zero, so that the field $X$ is at first minimally coupled to gravity.
\item Massive particles $X$, previously in thermal equilibrium with the SM plasma, decouple and quickly become non-relativistic, eventually forming atoms by standard recombination. In appendix \ref{recombination} we show that this is possible in a certain region of the parameter space. 
\item The scalar field $\psi$ undergoes a phase transition and acquires a non-zero expectation value, which sends the non-minimal coupling parameter close to its maximal value. The field $X$ is now non-minimally coupled to gravity.
\item The newly created atoms start absorbing primordial gravitons of the right frequency. Some fraction of them is ionized. After leaving a discernible imprint on the primordial gravitational spectrum, they decay to radiation. 
\end{enumerate}

The calculation of the exact shape of the absorption line is heavily model dependent and beyond the scope of this paper. We will thus limit ourselves to a couple of considerations. For one thing, in any concrete cosmological scenario, absorption lines will be broadened by the expansion of the universe, since gravitons will be absorbed at different times. The size of the broadening will depend on the strength of the gravitational coupling and the rapidity of the decay: gravitons will keep being absorbed until the number density of bound states decreases below a critical value, and the atomic gas is not dense enough to sustain gravitational absorption. 

\begin{figure}
    \centering
    \includegraphics[width=11.7cm]{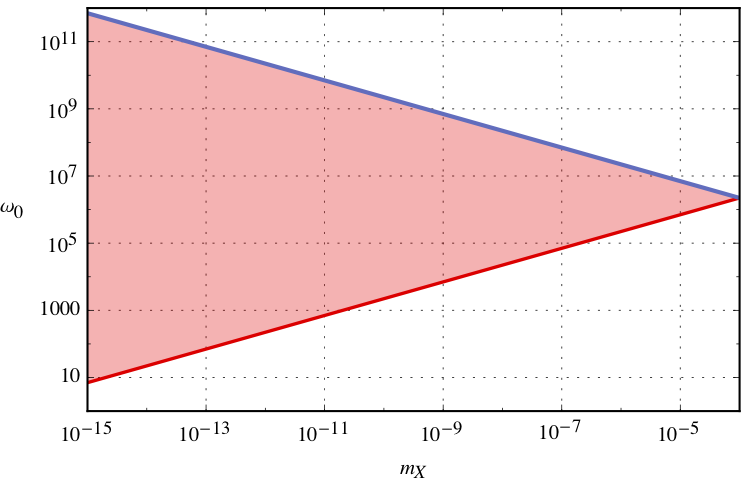}
    \caption{The shaded red region represents the peak frequency range $\omega_0$ for absorption lines if the universe is dominated by non-minimally coupled atoms with mass $m_X$, and $\alpha_X=\tilde{\xi}=0.1$. Absorption happens in the temperature range $10^3 m_X \sqrt{m_X/m_p}\lesssim T_{\text{abs}}\lesssim 0.1 \sqrt{m_X m_p}$. The red line corresponds to absorption at $ T_{\text{abs}} = 0.1 \sqrt{m_X m_p}$, while the blue line corresponds to absorption at $ T_{\text{abs}} =10^3 m_X \sqrt{m_X/m_p}$. The range of frequencies widen as the mass decreases. The mass is in Planck units, while the frequency is in Hertz. }
    \label{plot2}
\end{figure}

Secondly, the peak frequency of the absorption line will depend on both the binding energy of the atom and the expansion history of the universe. In the simplest scenario, in which gravitons are absorbed by atoms of mass $m_X$ and gauge charge $\alpha_X$ at temperature $T_{\text{abs}}$, and the universe evolves dominated by radiation from $T_{\text{abs}}$ until the present time, the peak frequency of the signal as measured today $\omega_0$ is 
\beq\label{omega0}
\omega_0 = \frac{2}{9} m_X \alpha_X^2 \frac{T_0}{T_{\text{abs}}},
\eeq
where $T_0$ is the average temperature of the universe today. Here we ignored the late stage of matter domination since it affects the final result only slightly. Fig.\ref{plot2} shows the range of frequencies where one could find absorption lines for a given atomic mass, assuming that absorption happens somewhere in the range given by (\ref{Hbound}), namely for $ 10^3 m_X \sqrt{m_X/m_p}\lesssim T_{\text{abs}}\lesssim 0.1 \sqrt{m_X m_p}$. The peak frequency is typically large and is of order $10^6$ Hz for strongly coupled atoms with the highest possible mass, $m_X \sim 10^{-4} m_p$, absorbing immediately after reheating ($T_{\text{abs}}\sim10^{-3} m_p$). Lighter atoms absorb at a lower frequency, but within a wider range of temperatures, leading to spectral lines from 10 to $10^{12}$ Hertz. 

Ionization of atoms will generically happen together with absorption. We compute the gravitational ionization cross section in the non-relativistic regime in appendix \ref{gravabs}. The result is
\beq
\sigma_{\text{ion}} = \frac{3 \times 2^{9} \pi^2}{5} \frac{\eta^6 (4+\eta^2)}{(1+ \eta^2)^4} \frac{e^{-4 \eta \cot^{-1}\eta}}{1-e^{-2 \pi \eta}} G,
\eeq
where $\eta=k_B/k$, $k_B$ is the Bohr momentum, and $k$ the final momentum of the ionized particle. The ionization cross section is maximal for $k=0$, and rapidly goes to zero for higher momenta. The maximum value is 
\beq
\sigma_{\text{ion,max}} = \frac{3 \times 2^{9} \pi^2}{5 e^4} G,
\eeq
which is about a hundred times bigger than (\ref{abssigma}). Therefore, with a flat spectrum, only a $10^{-2}$ fraction of the atoms will absorb, while the rest will be ionized by the gravitational radiation. We then expect discrete lines on top of some broad absorption feature. 

The intensity of the line can also vary significantly depending on the specific scenario. In particular, if the number of bound states greatly exceeds the number of gravitons at the absorbing frequency, the gravitons will be all be absorbed or rescattered, resulting in a near extinction of the signal at that frequency. Conversely, if there are more gravitons than bound states, the signal will only be partially dimmed. In our simple scenario we can directly compare the number density of bound states with the number density of primordial gravitons from inflation. The energy density spectrum of tensor modes from inflation is \cite{Caprini:2018mtu}
\beq\label{GWspectrum}
\Omega_{\text{GW}}(k)=\frac{3}{128} \Omega_{\text{rad}} \mathcal{P}_h (k) \left[ \frac{1}{2} \left(\frac{k_{\text{eq}}}{k}\right)^2+\frac{4}{9} (\sqrt{2}-1)\right],
\eeq
where $ \Omega_{\text{rad}}$ is the density parameter of radiation, $ k_{\text{eq}}$ is the wave number of modes that re-enter the horizon at matter-radiation equality, and $\mathcal{P}_h (k)$ the inflationary tensor power spectrum, given by
\beq
 \mathcal{P}_h (k)  \simeq \frac{2}{\pi^2} \frac{H_i^2}{m_p^2},
\eeq
where $H_i$ is the scale of inflation. The spectrum (\ref{GWspectrum}) is flat for modes that entered the horizon during the radiation era, and scales as $k^{-2}$ for modes that entered the horizon during the matter era. We are interested in absorption in the early universe, deep in the radiation dominated era, therefore in our case $k \gg k_{\text{eq}}$, and the spectrum is flat. The energy density in gravitational waves at frequency $k$ is just $\rho_{\text{GW},k}\simeq \rho_c \Omega_{\text{GW}}(k) =  \omega_k n_{\text{GW},k}$, where $\omega_k=k$ is the energy of a graviton of frequency $k$, $n_{\text{GW},k}$ the number density of gravitons at that frequency, and $\rho_c$ the critical density. The number density $n_{\text{GW},k}$ then is
\beq
n_{\text{GW},k} \approx \frac{\rho_{\text{rad}}}{k} \frac{H_i^2}{m_p^2}.
\eeq
On the other hand, if the bound states are close to saturating the critical energy density, as we assume, their number density is just 
\beq
n_B=\frac{\rho_\text{rad}}{m_X}. 
\eeq
The absorption frequency is roughly $k \approx \alpha_X^2 m_X$, therefore the bound states will dominate whenever $\alpha_X \gtrsim H_i/m_p$. Given that the current bound on the energy scale of inflation is $H_i \lesssim 10^{-6} m_p$, bound states will typically be more numerous than gravitons, and the resulting absorption lines quite sharp.  

Note also that although atoms interact more strongly with gauge forces than with gravity (even with a strong non-minimal coupling), photon absorption is irrelevant in this context. The typical frequency (energy) of CMB photons is given by the temperature of the background radiation $T$. In order for the atoms to be decoupled from the radiation bath, their binding energy should be higher than the temperature, $\alpha_X^2 m_X>T$, therefore CMB photons will generically not be energetic enough to be absorbed in atoms. The cosmic gravitational background, on the other hand, is assumed to contain gravitons of all frequencies, provided that their wavelength is contained inside the horizon, so it will also contain gravitons capable of exciting the atoms. In other words, while the CMB has a black body spectrum peaked at some frequency that is typically too low to excite atoms, the cosmic gravitational background is flat and contains gravitons of all frequencies, including the ones capable of exciting or ionizing the atoms. 

While we do not think that the very special type of non-minimal coupling we presented in this section represents a realistic scenario for graviton absorption, it illustrates the kind of new physics that one needs to have in order to produce absorption lines in gravitational spectra. In this particular case, the equivalence principle is violated on-shell by the particles making up the atom, leading to a stronger coupling to gravity. The non-minimal coupling also breaks the universality of gravity, thereby introducing other-than-gravitational scales in the cross section. 

Apart from being a strong hint to new physics beyond the standard model of cosmology and particle physics, detection of these absorption lines would provide direct evidence for the quantisation of the gravitational field, and the existence of gravitons. Absorption at a single frequency is only possible if the gravitational field is made up of quanta whose energy is determined by their frequency (Einstein's relation $E= \hbar \omega$). Experimental observation of gravitational absorption lines would then rule out all scenarios in which the gravitational field is a purely classical entity, like for example in models of entropic gravity. In the specific scenario we discussed in this section, however, where absorption lines arise in the primordial gravitational spectrum produced by inflation, the mere fact that a flat primordial spectrum is there might be enough evidence to deduce that the gravitational field is quantized, as argued in \cite{Krauss:2013pha}.

\section*{Acknowledgments}

We thank Niklas G. Nielsen for initial discussions. In addition, we would also like to thank Gia Dvali, Alex Kehagias, Florian Niedermann and Antonio Riotto for helpful comments. This work is supported by Villum Fonden grant 13384. CP3-Origins is partially funded by the Danish National Research Foundation, grant number DNRF90.

\newpage

\begin{appendix}
\numberwithin{equation}{section}
\setcounter{equation}{0}

\section{Graviton absorption and ionization}\label{gravabs}
We derive the absorption cross section (\ref{abssigma}) for hydrogen-like atoms in two different ways. We start with a heuristic classical argument leading to the planck area cross section, followed by a detailed QFT calculation that confirms the result. We also derive the general formula for the gravitational ionization cross section and discuss different limits of it. We then prove that the absorption cross section is always Planckian for bound states with arbitrary potential, using only the Schrodinger equation. Finally, we extend the conclusions to non-minimally coupled theories. 

\subsection{Absorption cross section: heuristic derivation}\label{heuristic}

This section is taken from \cite{Misner:1974qy}.

The simplest idealized graviton detector is an oscillator driven by a steady flux of gravitational waves. The oscillator consists of two point masses $m$ attached at the ends of a spring of length $L$, with a natural frequency of vibration $\omega_0$ and a damping time $\tau_0 \gg 1/\omega_0$. Its equation of motion is 
\beq
\frac{d^2 \xi}{dt^2}+\frac{1}{\tau_0} \frac{d\xi}{dt} + \omega_0^2 \xi = \frac{d^2 \xi}{dt^2}|_d,
\eeq
where $\xi$ is the displacement of the two masses and the term on the right is the driving acceleration due to the wave. A wave traveling in the z-direction past the detector in the transverse-traceless (TT) gauge can be written as
\begin{align}
h_{xx}^{TT} = - h_{yy}^{TT}=A_+(t-z) \nonumber \\
h_{xy}^{TT}=h_{yx}^{TT}=A_X (t-z),
\end{align}
where the amplitudes $A_+$ and $A_X$ represent the two independent modes of polarization. Let's suppose that the impinging wave has frequency $\omega$ and +  polarization ($A_X=0$) with $A_+=h e^{-i \omega (t-z)}$. We also assume that the detector is much smaller than the wavelength, so that one can set $z = 0$. Then, the tidal acceleration produced by the wave is 
\begin{align}\label{tidal}
 \frac{d^2 x}{dt^2}|_d= - R_{x0j0} x^j = -\frac{1}{2} \omega^2 h e^{-i \omega t} x \nonumber \\
 \frac{d^2 y}{dt^2}|_d= - R_{y0j0} x^j = +\frac{1}{2} \omega^2 h e^{-i \omega t} y,
\end{align}
where $R_{\mu \nu \alpha \beta}$ is the Riemann curvature tensor. Denoting with $\theta$ and $\phi$ the polar angles of the detector relative to the wave axes, the total driving acceleration is
\beq
 \frac{d^2 \xi}{dt^2}|_d= \frac{x}{L}  \frac{d^2 x}{dt^2}|_d +  \frac{y}{L}  \frac{d^2 y}{dt^2}|_d +  \frac{z}{L}  \frac{d^2 z}{dt^2}|_d=-\frac{1}{2} \omega^2 h L e^{-i \omega t} \sin^2(\theta) \cos(2 \phi),
\eeq
and the equation of motion for the oscillator gives
\beq
\frac{d^2 \xi}{dt^2}+\frac{1}{\tau_0} \frac{d\xi}{dt} + \omega_0^2 \xi =-\frac{1}{2} \omega^2 h L e^{-i \omega t} \sin^2(\theta) \cos(2 \phi),
\eeq 
with a steady state solution given by (it is understood that one should take the real value)
\beq
\xi(t)=\frac{\omega^2 h L \sin^2(\theta) \cos(2 \phi)}{2(\omega^2-\omega_0^2 +i\omega/\tau_0)} e^{-i \omega t}.
\eeq
When the incoming waves are near resonance with the detector own frequency, $(\omega - \omega_0) \lesssim 1/\tau_0 \ll \omega_0$ (assuming $\omega>0$), the solution becomes
\beq
\xi(t)=\frac{\omega_0 h L \sin^2(\theta) \cos(2 \phi)}{4(\omega-\omega_0 +\frac{i}{2 \tau_0} )} e^{-i \omega t}.
\eeq
Then, the time-averaged vibrational energy of the detector is
\beq
\langle E_v \rangle=2 \, \frac{1}{2}  m  \langle \dot{\xi}^2 \rangle =\frac{1}{16} \frac{m L^2 \omega_0^4 h^2 \sin^4(\theta) \cos^2(2 \phi)}{(\omega-\omega_0)^2+(1/2 \tau_0)^2}
\eeq
Gravitational wave production by the motion of the detector is negligible, therefore the energy dissipation rate $E_v/\tau_0$ can be equated to the rate at which the detector absorbs energy from the incoming waves, which is in turn equal to the (polarized) cross section for absorption $\sigma_{\text{abs},P}$ times the incoming flux:
\beq
E_v/\tau_0=  \frac{1}{32 \pi G} \sigma_{\text{abs},P} \omega^2 h^2.
\eeq
Consequently, near resonance, the polarized cross section for absorption of gravitational waves is
\beq
 \sigma_{\text{abs},P}= \frac{2 \pi G m L^2 (\omega_0^2/\tau_0)  \sin^4(\theta) \cos^2(2 \phi)}{(\omega-\omega_0)^2+(1/2 \tau_0)^2}.
\eeq
Averaging over all polarizations we obtain the unpolarized cross section, which is given by the Lorentzian
\beq
\tilde{ \sigma}_{\text{abs}}= \frac{(8 \pi/15) G m L^2 (\omega_0^2/\tau_0)}{(\omega-\omega_0)^2+(1/2 \tau_0)^2},
\eeq
and the cross section averaged over all frequencies is 
\beq
 \sigma_{\text{abs}}= \frac{1}{\omega_0} \int_{-\infty}^{+\infty}\tilde{ \sigma}_{\text{abs}} d\omega=\frac{16}{15} \pi^2 G m L^2 \omega_0.
\eeq
Therefore, the gravitational absorption cross section for a generic detector of mass $m$, size $L$, and proper frequency $\omega_0$ goes like $ \sigma_{\text{abs}} \sim GmL^2\omega_0$. If the detector is an atom, the standing-wave quantisation condition $mL^2\omega_0 = n \in \mathbb{N}$ places a further constraint on the parameters, and the cross section simply becomes $\sigma_{\text{abs}} \sim G = l_p^2$. We deduce that the Planck squared cross section is solely a result of angular momentum quantisation, and does not depend on the specifics of the atom. In particular, if the atom is bound by gravity \cite{Nielsen:2019izz} the quantisation condition still applies, with $L\equiv r_B=(m \alpha_G)^{-1}$ and $\alpha_G=m^2/m_p^2$. 

\subsection{Absorption cross section: QFT computation}

The previous derivation is heuristic at best, and merely gives an intuitive understanding of the way the planck area cross section emerges from classical gravity. The gist of it is that any oscillatory system with mass $m$, size $L$, and frequency $\omega$, absorbs gravitons with cross section $\sigma_{\text{abs}} \sim G m L^2 \omega$; atoms are particular oscillatory states with quantized angular momentum, $mL^2\omega_0 = n $, hence they absorb with a cross section proportional to $G$. The classical derivation however is not fully satisfactory and does not give the correct numerical result. For this reason, we now derive the exact cross section in a fully consistent way using field theoretic methods. This will also allow us to generalize the result to non-minimally coupled matter. In the following we will denote the mass and electromagnetic coupling of the atom with $m$ and $\alpha$ respectively.

In linearized gravity, for small metric deviations $h_{\mu\nu}=g_{\mu\nu}-\eta_{\mu\nu} \ll 1$, the interaction Lagrangian density is given by 
\beq
\mathcal{L}=\frac{1}{2} h_{\mu\nu} T^{\mu\nu},
\eeq
where $T_{\mu\nu}$ is the stress energy tensor of matter. The interaction Hamiltonian is given by $H=pv-L$, where $L$ is the interaction Lagrangian (Lagrangian density integrated over space), and $p$ and $v$ the momentum and velocity of the particle respectively. In a local inertial frame (LIF), the dominant term to the stress energy tensor is the mass-energy density, so $\mathcal{L} \approx \frac{1}{2} h_{00} T^{00}$. Moreover, the generalized velocities are negligible, therefore $H\approx-L=-\frac{1}{2} m h_{00}$, where $m$ is the localized mass of the system. 

In a LIF, the time-time component of the metric deviation can be written as \cite{Boughn:2006st}
\beq
h_{00} = - \frac{1}{2} \omega^2 h e^{i(q \cdot x-\omega t)} x^j x^k e_{jk} + \text{c.c.}, 
\eeq
where $h$, $q$, and $\omega$ are the amplitude, momentum and energy of the impinging gravitational wave, respectively. Hence, the interaction Hamiltonian is
\beq\label{hamiltonian}
H=\frac{1}{4} m \omega^2 h x^j x^k e^{i(q \cdot x-\omega t )} e_{jk}  + \text{c.c.}
\eeq
To simplify matters, we assume that the atom interacts with a single graviton. Then, the amplitude $h$ is simply
\beq
h = \sqrt{8 \pi G} \omega.
\eeq

We work in the dipole approximation, namely we assume that the wavelength of the gravitational wave (graviton, in our case) is much larger than the extent of the atom, so that $q r_B \ll 1$ and $e^{iq \cdot x} \approx 1$. In first order perturbation theory, the transition probability per unit time between two atomic states $\Psi_{1s}$ and $\Psi_{3d2}$ is equal to (Fermi's Golden Rule)
\beq
\Gamma = \frac{2 \pi}{\omega} |\Braket{\Psi_{3d2} | H |\Psi_{1s}}|^2  = \frac{2 \pi^2 G\omega^5}{5} \left( D_{ij}^*D^{ij}-\frac{1}{3} |D^i_i|^2 \right),
\eeq
where
\beq\label{quadrupole}
D_{ij} = m \int \Psi_{3d2}^* x_i x_j \Psi_{1s} d^3 r
\eeq
is the mass quadrupole tensor, and the average is taken over all directions of the incident gravitational wave. 
When a graviton is absorbed, the transition occurs between the 1s and the 3d2 states, whose normalized wavefunctions are
\beq\label{wavefunctions}
\Psi_{1s} = \frac{1}{\sqrt{\pi} r_B^{3/2}} e^{-r/r_B} \; \; ; \; \; \Psi_{3d2} = \frac{1}{162 \sqrt{\pi}} \frac{1}{r_B^{3/2}} \left( \frac{r^2}{r_B^2}\right) e^{-r/3r_B} \sin^2\theta e^{2 i\phi}.
\eeq 

The quadrupole components for this transition process are
\bea
D_{zz}=D_{xz}=D_{yz}=0 \\ 
D_{xx}=-D_{yy} =  i D_{xy} = \frac{3^4 }{2^8} m r_B^2,
\eea
with $r_B$ the Bohr radius of the atom. Finally, the absorption rate for the $1s \rightarrow 3d2$ transition is 
\beq
\Gamma = \frac{3^8 \pi^2}{5 \times 2^{13}} G m^2 r_B^4 \omega^5 .
\eeq
This gives the transition rate between the 1s and the 3d2 states when the atom is hit by a graviton of frequency $\omega$. The absorption cross section is just $\sigma_{\text{abs}}=\Gamma/\omega^3$, 
\beq
\sigma_{\text{abs}} =  \frac{3^8 \pi^2}{5 \times 2^{13}} G m^2 r_B^4 \omega^2 .
\eeq
For a 2-particle atom, $\omega = (4/9) \alpha^2 m$, and $r_B = (\alpha m)^{-1}$, so
\beq\label{A30}
\sigma_{\text{abs}} =  \frac{3^4 \pi^2}{5 \times 2^9} \, G \approx 0.31 \, l_p^2 . 
\eeq
The planck area cross section is retrieved in a full QFT calculation, with the correct numerical prefactor. 

\subsection{Ionization cross section}

One can use the same machinery to compute the ionization cross section. As before, we need to evaluate the matrix element $\Braket{\Psi_{f} | H |\Psi_{i}}$ between the initial hydrogenic ground state $\Psi_i$, and a plane wave final state $\Psi_f$, with
\beq\label{planewave}
\Psi_{i} = \frac{1}{\sqrt{\pi} r_B^{3/2}} e^{-r/r_B} \; \; ; \; \; \Psi_{f} = \frac{1}{L^{3/2}} e^{i k \cdot r},
\eeq
where we normalize the plane wave in a box of dimension $L$, and $k$ is the final momentum of the emerging ionized particle. The final momentum $k$ satisfies $k \ll m$, since we are working in the non-relativistic regime. Moreover, the plane wave solution for the final state is only valid when the graviton energy is much larger than the binding energy of the atom (Born approximation). For this reason, the final result holds in the regime $k_B \ll k \ll m$.

Fermi's Golden Rule gives
\beq\label{golden}
\Gamma = 2 \pi \rho(k) |\Braket{\Psi_{f} | H |\Psi_{i}}|^2,
\eeq
where $\rho(k)$ is the density of final states, given by
\beq
\rho(k) = \frac{m k L^3}{2 \pi^2}.
\eeq
For an incident graviton of amplitude $h=\sqrt{8 \pi G} \omega$, the transition rate is
\beq\label{TRI}
\Gamma = \frac{3 \times 2^{11} \pi}{5} \frac{h^2 \omega^4 m^3 r_B^{11} k^5}{(1+r_B^2k^2)^8}.
\eeq
The incident graviton energy $\omega$ is equal to the sum of the binding energy and the kinetic energy of the emerging particle, namely 
\beq\label{energycons}
\omega=\alpha/2 r_B + k^2/2m.
\eeq
Using (\ref{TRI}) and (\ref{energycons}), we obtain the gravitational cross section for ionization $\sigma_{\text{ion}}=\Gamma/\omega^3$ in the high energy regime,
\beq\label{sigmaion}
\sigma_{\text{ion}}=  \frac{3 \times 2^{10} \pi}{5} \frac{ (r_B k )^5}{(1+r_B^2k^2)^5} G.
\eeq
This cross section is always much smaller than the planck area in its regime of validity.

It is possible to compute the ionization cross section in the non-relativistic limit ($k \ll m$) without resorting to the Born approximation, and therefore extend the result also for small final momenta. The computation was first carried out (to our knowledge) in \cite{Burundukov:1985cd}. The continous-spectrum wavefunction for scattering in a Coulomb field which asymptotes to a box-normalized plane wave $\sim L^{-3/2} \exp(i k\cdot r)$ in the non-relativistic limit is given by\cite{Berestetsky:1982aq}
\beq\label{continouswave}
\Psi_{f} = \frac{1}{L^{3/2}} \exp\left(i k \cdot r+\frac{\pi}{2} \eta \right) \Gamma(1- i \eta) \mathcal{F}\left[-i \eta, 1, -i(k\cdot r+ k r)\right],
\eeq
where $\Gamma$ is Euler's gamma function, $\mathcal{F}[a,b,c]$ denotes Kummer's confluent hypergeometric function, and $\eta \equiv \alpha/v$, where
\beq
v=\sqrt{\frac{2}{m}(\omega-\alpha/2 r_B)} = \frac{k}{m},
\eeq
is the final velocity of the ionized particle. In (\ref{continouswave}), $k \cdot r$ represents the 3-vector inner product, while $k r$ is the simple product between the vector magnitudes. Therefore, if we imagine the particle to be ejected in the z-direction, $(k\cdot r+ k r)= kr(1+\cos(\theta))$. For large velocities $\eta \rightarrow 0$, (\ref{continouswave}) reduces to the plane wave solution of (\ref{planewave}). Plugging (\ref{continouswave}) into (\ref{golden}) we find the following cross section,
\beq\label{finalion}
\sigma_{\text{ion}} = \frac{3 \times 2^{9} \pi^2}{5} \frac{\eta^6 (4+\eta^2)}{(1+ \eta^2)^4} \frac{e^{-4 \eta \cot^{-1}\eta}}{1-e^{-2 \pi \eta}} G.
\eeq
The result is reminiscent of the photoionization cross section, although the two differ crucially in the $\eta$ dependence. This cross section is valid for all $k\ll m$. In particular, one can retrieve (\ref{sigmaion}) in the high energy limit $\eta \rightarrow 0$: the ionization cross section for high momenta falls off as $(k/k_B)^{-5}$. In the opposite limit, $\eta \rightarrow \infty$, the cross section approaches a constant value,
\beq\label{sigmaion2}
\sigma_{\text{ion}} \xrightarrow{\eta \rightarrow \infty} \frac{3 \times 2^{9} \pi^2}{5 e^4} G,
\eeq
which is about 50 times larger than $G$. The gravitational ionization cross section of hydrogen was also computed in \cite{Smolin:1982jt}, but with a different result. The author found a cross section that vanishes for $k=0$, and is proportional to $k/k_B$ in the low energy limit. 

The ionization cross section (\ref{finalion}) is always of order $G$ or smaller. It reaches a maximum at $k=0$, and rapidly goes to zero for $k/k_B \gg 1$.  

\subsection{Multi-particle atoms and generic bound states}
The absorption cross section in (\ref{A30}) is technically only valid for 2-particle atoms. Is the cross section of the order of the planck area also for multi-particle atoms? The classical computation gives us a clue. There, we saw that the planck area emerged merely as a result of angular momentum quantization, and the angular momentum of every particle in an atom needs to be quantized simply because of standing wave considerations. 

Concretely, take a multi-particle atom with $N>2$ particles. The details of the atomic structure do not matter that much, and will not affect the final result. The Schrodinger equation can only be solved exactly in the case of two-particle atoms; the orbitals of multi-particle atoms are found by methods of iterative approximation. However, orbitals of multi-particle atoms are qualitatively similar to those of hydrogen, and in first approximation, they can be taken to have the same form. The total wavefunction of the whole atom is then just a direct product of single particle hydrogen-like atomic orbitals. The particles in the outer orbit are typically the ones responsible for the absorption by transitioning to a higher energy state. In the atomic orbital approximation, they feel a potential $Z^* \alpha /r$, where $Z^*$ is the effective charge due to the inner particles. The energy levels therefore are
\beq
E_n = -m \frac{ (Z^* \alpha)^2}{4 n^2}.
\eeq

For a given principal quantum number $n$, and angular momentum quantum number $l$, the wavefunction is proportional to
\beq
\Psi_{n,l} \propto  \left(\frac{Z^*}{r_B}\right)^{3/2} \left(\frac{Z^* r}{r_B}\right)^l \exp\left({-\frac{Z^*r}{n r_B}}\right).
\eeq 
We can then compute the mass quadrupole tensor (\ref{quadrupole}) for the transition $(n_1,l) \rightarrow (n_2,l+2)$ and plug it in the formula for the cross section. Schematically, the non-zero components of the quadrupole tensor scale like
\beq
D_{ij} \propto m \frac{r_B^2}{(Z^*)^2},
\eeq
thus the absorption cross section is 
\beq
\sigma_{\text{abs}} \propto  G \omega^2 |\langle D_{ij} \rangle|^2 \sim  G, 
\eeq
given that $\omega=E_{n_2}-E_{n_1} \sim m (Z^* \alpha)^2 $, and $r_B = (\alpha m)^{-1}$. Intuitively, this has to do with the fact that for a multi-particle atom the Bohr radius $r_B$ is rescaled by $1/Z^*$, while the frequency $\omega$ changes by $(Z^*)^2$, therefore the product $m^2 \omega^2 r_B^4$ is independent of $Z^*$. Multi-particle atoms will absorb approximately with the same probability as two-particle atoms. 

We can also ask whether the result can be extended to other types of bound states. For example, is it true also for a neutron or a proton bound in a nuclear potential? The simplest model of the atomic nucleus is the nuclear shell model. The nuclear potential is well approximated by the three dimensional harmonic oscillator, plus a spin-orbit interaction that we can neglect:
\beq
V(r) = \frac{1}{2} m \omega^2 r^2.
\eeq
Here $r$ is the distance between the nucleons, $m$ their mass and $\omega$ controls the strength of the interaction. The energy levels are
\beq
E_{n,l} = \omega\left( n + l+ \frac{3}{2} \right),
\eeq
where $n$ is the radial quantum number and $l$ the angular momentum quantum number. Schematically, the corresponding nuclear wavefunctions are
\beq
\Psi_{n,l} \propto R^{-3/2} \left( \frac{r}{R}\right)^{n+l}\exp\left({-\frac{r^2}{2R^2}}\right),
\eeq
where $R=(m \omega)^{-1/2}$ is the typical size of the nucleus. The non-zero components of the mass quadrupole tensor for the transition $(n_1,l) \rightarrow (n_2,l+2)$ are
\beq
D_{ij} \propto m \frac{R^2}{2} (5+2l+n_1+n_2),
\eeq
therefore, since $\omega=E_{n_2,l+2}-E_{n_1,l}$ and $R = (m\omega)^{-1/2}$, the absorption cross section is
\beq
\sigma_{\text{abs}} \propto G m^2 \omega^2 R^4 \sim G.
\eeq

It is no coincidence that we found $\sigma_{\text{abs}} \sim l_p^2$ for atoms and nuclei alike, as the result is much more general and actually applies to all bound states, as we will show now. The non-relativistic Schrodinger equation for a particle in a potential $V(r)$ is
\beq
\left[-\frac{1}{2m} \nabla^2 + V \right] \Psi = E \Psi.
\eeq
A bound state is defined as a quantum state for which $E<V(+\infty)$. The solutions of the Schrodinger equation with $E<V(+\infty)$ have the property that the wavefunction $\Psi$ rapidly goes to zero at large distances, which means that the particle is confined to a region of space. We need to examine separately the cases in which $V(\infty)$ is finite and infinite. In the former case, we can just add a constant to the potential to make it zero at infinity, so the condition on the total energy becomes $E<0$. The Schrodinger equation at large distances becomes
\beq\label{schrod}
\frac{1}{2m} \nabla^2{\Psi} \approx  |E| \Psi,
\eeq 
therefore 
\beq
\Psi(r) \xrightarrow{r \rightarrow \infty} \exp \left( - \sqrt{2 m |E|} r\right) \equiv \exp \left( -2  r/R\right),
\eeq
where
\beq\label{sizeBS}
R= \frac{1}{\sqrt{m |E|}}
\eeq
can be interpreted as the size of the bound state. 

If $V(r)$ diverges at infinity, like in the case of the harmonic oscillator, then the wavefunction will go to zero even faster, and (\ref{sizeBS}) will give an upper bound on the size of the bound state. This is because in equation (\ref{schrod}), $|E|$ would be replaced by $V(r)$, which is a monotically increasing function in that limit. Moreover, the case in which $V(r) \xrightarrow{r \rightarrow \infty} \infty$ is not realistic, since the potential cannot diverge in any physical system and it will always reach a plateu. For example, the nuclear potential at large distances is better approximated by the Woods-Saxon potential $V(r)= - V_0/(1+\exp(r/R))$, which approaches zero at large distances. 

The upshot is that any quantum bound state has a size given roughly by (\ref{sizeBS}), $R=(m |E|)^{-1/2}$, where $E$ is the energy of the quantum state. The wavefunction at large distances is just a polynomial times a decreasing exponential, therefore the mass quadrupole moment (\ref{quadrupole}) $D_{ij}$ is always proportional to $m R^2$. Thus, the absorption cross section is 
\beq
\sigma_{\text{abs}} \sim G m^2 |E|^2 R^4,
\eeq
which, given (\ref{sizeBS}), is naturally Planckian. We conclude that every quantum bound state absorbs with a planck area cross section. 

\subsection{Planck suppressed corrections}

Planck-suppressed operators can affect the final result. Take for example the case in which the matter in the atom is non-minimally coupled through the term 
\beq\label{Lxi}
\mathcal{L}_\xi = \frac{\xi}{m_p^4} R_{\mu \nu \alpha \beta} \partial^\mu \partial^\alpha X \partial^\nu \partial^\beta X .
\eeq
We choose to work in the transverse-traceless (TT) gauge. In this gauge, the metric perturbations satisfy $\partial_\mu h^\mu_\nu = 1/2 \, \partial_\nu h^\mu_\mu$, and $h_{\mu 0}=h^\mu_\mu =0$. A metric perturbation with amplitude $h$ and polarization tensor $e_{\mu\nu}$ can be written as $h_{\mu \nu} = h e_{\mu \nu}$, with $e_{00}=e_{\mu 0}=e^\mu_\mu=0$. Consequently, a harmonic plane gravitational wave in the TT gauge can be written as
\beq
h_{ij}= h e^{i (q \cdot x-\omega t)} e_{jk} + c.c.
\eeq
where $\omega$ and $q$ are the energy and momentum of the wave. To lowest order in $h_{\mu\nu}$, the Riemann curvature tensor $R_{\mu\nu\alpha\beta}$ is
\beq
R_{\mu \nu \alpha \beta} = \frac{1}{2} \left( \partial_\nu \partial_\alpha h_{\mu \beta} + \partial_\mu \partial_\beta h_{\nu \alpha} -  \partial_\mu \partial_\alpha h_{\nu \beta}  -  \partial_\nu \partial_\beta h_{\mu \alpha}  \right).
\eeq
In the linear theory, $R_{\mu \nu \alpha \beta}$ is invariant under gauge transformations $x^\mu \to {x'}^\mu = x^\mu -\xi^\mu$, since $h_{\mu\nu}$ transforms as $h_{\mu\nu} \to {h'}_{\mu\nu}= h_{\mu\nu}-\partial_\mu\xi_\nu -\partial_\nu\xi_\mu$. In the non-relativistic limit, the dominant contribution to (\ref{Lxi}) is given by $ (\xi/m_p^4) R_{0 i 0 j} \partial^0 \partial^0 X \partial^i \partial^j X$, where (in TT gauge)
\beq
R_{0 i 0 j}=\frac{1}{2} \omega^2 h e^{i (q \cdot x-\omega t)} e_{jk} +\text{ c.c.}
\eeq
Since $R_{0 i 0 j}$ is gauge invariant, it will take this value also in a locally inertial frame (LIF). In a LIF, $\partial^0 \partial^0 X \sim m$ and $ \partial^i \partial^j  X \sim k^i k^j/m$, where $m$ and $k$ are the mass and the momentum of the particles in the bound state. As before, we work in the dipole approximation, $e^{i q \cdot x} \approx 1$.

Consequently, the interaction Hamiltonian is
\beq\label{HHxi}
H_\xi = \frac{1}{4} \frac{\xi}{m_p^4} m \omega^2 h \, k^i k^j \, e^{-i\omega t} e_{ij} + \text{c.c.}
\eeq
This differs from (\ref{hamiltonian}) simply by the replacement $x^i x^j \rightarrow k^i k^j$. Following the same steps as before, Fermi's Golden Rule for the absorption of a single graviton gives
\beq
\Gamma =  \frac{2 \pi}{\omega}|\Braket{\Psi_{3d2} | H |\Psi_{1s}}|^2 =  \frac{2 \pi^2 G\omega^5}{5} \frac{\xi^2}{m_p^8} \left( \tilde{D}_{ij}^*\tilde{D}^{ij}-\frac{1}{3} |\tilde{D}^i_i|^2 \right),
\eeq
where now
\beq
\tilde{D}_{ij} = m \int \Psi_{3d2}^* (k) k_i k_j \Psi_{1s} (k) d^3 k,
\eeq
and $\Psi_{1s}(k)$ and $\Psi_{3d2}(k)$ are the momentum-space wavefunctions
\beq
\Psi_{1s}(k) = r_B^{3/2} \frac{2 \sqrt{2}}{\pi} \frac{1}{(k^2r_B^2+1)^2} \; \; ; \; \; \Psi_{3d2}(k) = r_B^{3/2} \frac{2^4\times 3^3 \sqrt{3}}{\pi} \frac{k^2 r_B^2}{(9 k^2 r_B^2+1)^4} (3 \cos^2(\theta)-1).
\eeq

The components of the tensor $\tilde{D}_{ij}$ are
\bea
\tilde{D}_{zy}=\tilde{D}_{xy}=\tilde{D}_{xz}=0 \\ 
\tilde{D}_{zz}=-\tilde{D}_{xx} =  - \tilde{D}_{yy} = \frac{1}{10 \sqrt{6}} \frac{m}{r_B^2},
\eea
and the absorption cross section is
\beq\label{sigmaabsxi}
\sigma_{\text{abs},\xi} = \frac{2 \pi^2}{3^4 \times 5^3}\frac{\xi^2 G}{m_p^8}  k_B^8,
\eeq
where $k_B = \alpha m$ is the Bohr momentum of the particles in the atom. 

Then, the absorption cross section with the non-minimal contribution is
\beq\label{sigmaxi}
\sigma_{\text{abs}} = \frac{3^4 \pi^2}{5 \times 2^9}  \, G \left[1 +\frac{2^{10}}{3^8 \times 5^2} \xi^2 \left( \frac{m}{m_p}\right)^8 \alpha^{8} \right].
\eeq

\begin{comment}
We can check the result by computing the amplitude for the absorption process in Fig.\ref{absorption} and comparing the cross sections. The amplitude squared is 
\begin{figure}
    \centering
    \includegraphics[width=8cm]{absorption.jpg}
    \caption{This diagram describes the absorption of a graviton by an X particle, as well as the time-reversed process in which the graviton is emitted (gravitational bremsstrahlung).}
    \label{absorption}
\end{figure}
\beq
| \mathcal{M}_A |^2 \sim \frac{ \xi^2}{m_p^{10}}  s^2 (s-4 m_X^2)^4 ,
\eeq
where $s\approx (2m_X+k^2/ m_X)^2$ is the center of mass energy squared of the process and $k$ the momentum of the particle. The correction to the absorption cross section then is 
\beq
\delta \sigma_{\text{abs}} \sim \frac{\xi^2}{m_p^{10}} (s-4 m_X^2)^4 \sim  \frac{\xi^2}{m_p^{10}}k_B^8,
\eeq
in agreement with (\ref{sigmaxi}), since $k_B=\alpha_X m_X$. Here we used the non-relativistic approximation $s\approx 4 (m_X^2 + k^2)$. That both the amplitude and the cross section vanish when the momentum is zero is easily explained by the fact that (\ref{Lxi}) is a derivative coupling, therefore it is not present when the particles are not moving. 
\end{comment}

The correction can become larger than one without violating the unitarity bound. By simple power counting, the scattering amplitude of $X$ particles interacting non-minimally through the coupling (\ref{Lxi}) goes like $\mathcal{M} \sim \xi \, E_X^{10}/m_p^{10}$, $E_X$ being the typical energy of the process. Specifically, for scattering in the s-channel, the amplitude squared is
\beq\label{MSs}
| \mathcal{M}_s |^2 = \frac{\xi ^4}{2^{12} m_p^{20}} s^6 \left(16 m^4-8 m^2 (s+4 t)+s^2+8 s t+8 t^2\right)^2,
\eeq
while for scattering in the t-channel, it is
\beq\label{MSt}
| \mathcal{M}_t |^2 = \frac{\xi ^4}{2^{12} m_p^{20}} t^6 \left(16 m^4-8 m^2 (4 s+t)+8 s^2+8 s t+t^2\right)^2,
\eeq
with $s\approx (2m+k^2/m)^2$ the center-of-mass energy squared and $t\approx 2 k^2 \ll s$ (non-relativistic regime). The unitarity bounds $| \mathcal{M}_{s,t} |^2 \lesssim 1$ then read
\beq\label{unitaritybound}
\xi \lesssim \frac{m_p^5}{m^3 k^2} \; \text{s-channel}  \; \; ; \; \; \xi \lesssim \frac{m_p^5}{m^2 k^3} \; \text{t-channel}. 
\eeq
\begin{comment}
\begin{figure}
    \centering
    \includegraphics[width=10cm]{exchange.jpg}
    \caption{X particles scattering via the non-minimal coupling to gravity (\ref{Lxi}).}
    \label{exchange}
\end{figure}
\end{comment}
Due to the high powers involved, numerical factors do not alter the bound significantly. Since $k \ll m$, the most restrictive of the two is the bound coming from s-channel scattering, $\xi \lesssim m_p^5/m^3 k^2$.  In the limit in which the non-minimal coupling dominates, we can rewrite the cross section using a rescaled coupling $\tilde{\xi} \equiv \xi/\xi_{\text{max}}$, with $\xi_{\text{max}} = m_p^5/(m^3 k_B^2)$. The cross section becomes
\beq
\sigma_{\text{abs}} =  \tilde{\xi}^2 \frac{ k_B^4}{m^6}=\tilde{\xi}^2 \frac{ \alpha^4}{m^2}.
\eeq 
$\xi=\xi_{\text{max}}$ gives the maximum possible cross section $\sigma_{\text{abs}} \sim \alpha^4/m^2$, which can be made arbitrarily large by taking large coupling and small masses.  

The amplitude for production of $X$ particles by SM particles in the s-channel, in the limit in which the kinetic energy is much larger than the mass, is 
\beq
|\mathcal{M}_{SM \rightarrow X}|^2 \simeq \frac{\xi ^2}{2^8 m_p^{12}} s^2(s+2t)^4 \sim \xi^2 \left(\frac{E}{m_p}\right)^{12},
\eeq
where $E$ is the typical kinetic energy of the incoming particles. The weakest bound on the non-minimal coupling from particle collider searches then just comes from requiring that the amplitude is less than one for $E \sim 10 \, \text{TeV} \sim 10^{-15} m_p$, which gives $\xi \lesssim 10^{90}$. In particular, $\xi_{\text{max}} \lesssim 10^{90}$, which means that the mass has to be at least greater than $m \gtrsim 10 \, \text{GeV}$ (since in the non-relativistic limit $k \ll m$). We stress that this is a very weak bound, and the actual bound from collider searches is likely to be much stronger. In any case, the bound cannot be stronger than $m \gtrsim 10 \, \text{TeV}$, due to the collision energy threshold at LHC. 

Note that the operator (\ref{Lxi}) breaks the equivalence principle, and as such it introduces a new additional force between the $X$ particles. This is the ultimate reason why the absorption cross section is no longer Planckian. To see this, let's look at the Hamiltonian for absorption of a graviton by an atom in minimally coupled gravity. In the non-relativistic limit, this is given by equation (\ref{hamiltonian}).
Classically, we can identify the Hamiltonian with the gravitational potential between the masses. In the heuristic derivation of section \ref{heuristic}, we consider an idealized graviton detector consisting of an oscillator driven by an external (gravitational) force. The driving acceleration due to the gravitational wave is given by equation (\ref{tidal}).
One can derive the classical driving acceleration also from the gradient of the Hamiltonian (\ref{hamiltonian}) as
\begin{equation}
    \frac{d^2 x_i}{dt^2}|_d = - \frac{1}{m} \partial_i H
\end{equation}
Using the driven harmonic oscillator equation 
\begin{equation}
\frac{d^2 \xi}{dt^2}+\frac{1}{\tau_0} \frac{d\xi}{dt} + \omega_0^2 \xi = \frac{d^2 \xi}{dt^2}|_d,
\end{equation}
one can then derive the planck area cross section as we have shown.

We can do something similar when the non-minimal coupling is present. The Hamiltonian now is given by equation (\ref{HHxi}).
The classical driving force of the oscillator then is
\begin{equation}
    F_\xi|_d = - \partial_i H_\xi = -\frac{1}{2} \frac{\xi}{m_p^4} m \omega^2 h k^2 k_i e^{-i\omega t}.
\end{equation}
This is an extra force that arises from the breaking of the equivalence principle on-shell. Note that $k$ here is the momentum of the particle inside the atom and is therefore of order $k_B$, the Bohr momentum. The classical driving force with minimal coupling instead is 
\begin{equation}
    F|_d = - \partial_i H = -\frac{1}{2} m \omega^2 h e^{-i \omega t} x _i.
\end{equation}

From the heuristic derivation of section \ref{heuristic}, it is clear that the absorption cross section is proportional to the square of the driving force, therefore the ratio between the cross section with non-minimal coupling and with minimal coupling is just (here we neglect numerical factors, as we are only interested in the general behaviour) 
\begin{equation}
    \frac{\sigma_{\text{abs},\xi}}{\sigma_{\text{abs}}} = \left( \frac{F_\xi|_d}{F|_d}\right)^2 = \frac{\xi^2}{m_p^8} k_B^8. 
\end{equation}
This agrees with the expression for the absorption cross section in equation (\ref{sigmaabsxi}).

\section{Recombination in the dark sector}\label{recombination}
Here we show that recombination is possible in the dark sector even at high energy scales, provided that a large chemical potential is present.

First of all, we assume that $X$ binds to another massive particle in the dark sector that we call $Y$. These particles are (oppositely) charged under a $U(1)$ gauge field, of similar mass $m_Y \approx m_X$, and they are both initially in thermal equilibrium. The reason why we don't consider the simpler scenario in which atoms are $X \bar{X}$ bound states is that, as we will see below, successful recombination at high energy scales requires a large chemical potential, i.e. a large matter-antimatter asymmetry in the dark sector. The following discussion is basically a retelling of the standard recombination of Hydrogen, just at a higher energy scale. 

The binding energy of $B=X Y$ bound states is
\begin{equation}\label{EB}
    E_B = \frac{\alpha_X^2 m_X}{4}.
\end{equation}
Assuming thermal equilibrium between the different species, we can write down the Saha equation for the recombination process:
\begin{equation}
    \frac{n_B}{n_X n_Y} = \left( \frac{2 \pi m_B}{m_X m_Y  T}\right)^{3/2} e^{E_B/T}.
\end{equation}
Electrical neutrality requires $n_X=n_Y$, thus
\begin{equation}
    \frac{n_B}{n_X^2}=\left( \frac{4 \pi}{m_X T}\right)^{3/2} e^{E_B/T}.
\end{equation}

We define the ionization fraction $F_X$ as
\begin{equation}
    F_X = \frac{n_X}{n_X + n_B}.
\end{equation}
Thus, we have 
\begin{equation}
    \frac{1 - F_X}{F_X^2} = \frac{n_B}{n_X^2} (n_X+n_B).
\end{equation}
At energies well below $m_X$, the number densities of $X$ and $Y$ particles are not exponentially decaying anymore but are determined by the matter-antimatter asymmetry in the dark sector, which we assume to be present. Therefore Saha equation becomes
\begin{equation}\label{saha}
     \frac{1 - F_X}{F_X^2} = \eta_X \frac{2 \zeta(3)}{\pi^2}  \left( \frac{4 \pi T}{m_X}\right)^{3/2} e^{E_B/T},
\end{equation}
where 
\begin{equation}
    \eta_X = \frac{n_X+n_B}{n_\gamma}
\end{equation}
is the ratio of X particles (free and bound) to photons in the universe. For baryons, this ratio is of order $\sim 10^{-9}$ (determined by the non-zero baryon number in our universe), but since we are considering particles in a dark sector that decay in the early universe, this number can be in principle larger.

We define the recombination temperature $T_{\text{rec}}$ as the temperature when $F_X=0.1$, so when 90\% of X particles are recombined. Solving for the temperature in (\ref{saha}) we obtain
\begin{equation}
    T_{\text{rec}} = - \frac{2}{3} \frac{E_B}{\text{ProductLog}[-\frac{2}{3} \eta_X^{2/3} A^{2/3} \frac{E_B}{m_X}]},
\end{equation}
where 
\begin{equation}
    A \equiv \frac{F_X^2}{1-F_X} \frac{2 \zeta(3)}{\pi^2}  (4 \pi)^{3/2} \approx 0.1.
\end{equation}
The argument of the ProductLog function is small and negative, so we can use the approximation $\text{ProductLog}(x) \approx \log(-x)$ for $x\rightarrow 0^-$. This gives us
\begin{equation}\label{Trec}
    T_{\text{rec}} \approx - \frac{2}{3} \frac{E_B}{\log(\frac{2}{3} \eta_X^{2/3} A^{2/3} \frac{E_B}{m_X})}.
\end{equation}
For the usual recombination of hydrogen atoms from electrons and protons, the formula above gives an approximate temperature of $0.3$ eV. In the paper we consider the case in which $\alpha_X \sim 0.1$. The mass range for obtaining absorption lines goes from $10^{-15} m_p$ to $10^{-4} m_p$. Taking a value of $m_X \sim 10^{-10} m_p$ and $\eta_X \approx 1$, for example, gives a recombination temperature of 
\begin{equation}
    T_{\text{rec}} \approx 7 \times 10^{-14} m_p,
\end{equation}
slightly below the binding energy of the atom (\ref{EB}) $E_B = \alpha_X^2 m_X/2 \sim 10^{-12} m_p$. As it turns out, the precise value of $\eta_X$ has a minor impact on the recombination temperature. This is all very similar to the standard recombination of hydrogen atoms. Notice that recombination takes place in an early matter-dominated era since $ T_{\text{rec}} < m_X$.

To derive equation (\ref{Trec}), we have assumed thermal equilibrium between the particles, therefore we need to make sure that decoupling of X particles from the plasma happens after recombination. The interaction rate of X particles with the plasma is
\begin{equation}
    \Gamma = n_X \sigma_T = \eta_X \sigma_T \frac{2 \zeta(3)}{\pi^2} T^3 F_X,
\end{equation}
where $\sigma_T=\alpha_X^2/m_X^2$ is the cross section for Thomson scattering. Decoupling occurs when $\Gamma \approx H$. In a matter-dominated phase 
\begin{equation}
    H^2= \frac{8 \pi G}{3} m_X n_X,
\end{equation}
therefore decoupling happens when
\begin{equation}\label{Td3}
    T_d^3 = \frac{8 \pi^3}{6 \zeta(3)} \frac{m_X^5}{m_p^2} \frac{1}{\alpha_X^4 \eta_X F_X(T_d)}.
\end{equation}
We can use the Saha equation (\ref{saha}) to deduce the temperature dependence of $F_X$. Substituting in (\ref{Td3}) and solving for the decoupling temperature we obtain
\begin{equation}
   T_d = \frac{2 }{9} \frac{E_B}{ \text{ProductLog}\left(\frac{E_B^9 m_p^8 \alpha_X^{16}\eta_X^2 \zeta(3)^2 }{2^4 3^{14} m_X^{17} \pi^{11}} \right)^{1/9} },
\end{equation}
which, in the limit $m_X \ll m_p$, becomes
\begin{equation}
   T_d = \frac{2 E_B}{ \log \left(B \eta_X^2 \frac{E_B^{17} m_p^8}{m_X^{25}} \right)},
\end{equation}
where 
\begin{equation}
B \equiv \frac{\zeta(3)^2 }{2^4 3^{14} \pi^{11}} \approx  6.4 \times 10^{-14}. 
\end{equation}
Recombination is efficient when $T_d < T_{\text{rec}}$, so that the vast majority of X particles is in bound states when decoupling occurs. The resulting constraint on the parameter space is
\begin{equation}
 \eta_X > \frac{1}{(A^2 B)^{1/4}} \alpha_X^{-10} \left( \frac{m_X}{m_p}\right)^2 \approx 10^{-4} \alpha_X^{-10} \left( \frac{m_X}{m_p}\right)^2.
\end{equation}
Assuming conservatively that $ \eta_X <1$ we get the final constraint on the mass as
\begin{equation}
m_X < 100 \alpha_X^5 m_p.
\end{equation}
In our model we assume $\alpha_X \sim 0.1$, therefore the constraint is $m_X < 10^{-3} m_p$, which is easily satisfied. Our parameter space allows for successful recombination, provided that $\eta_X$ is sufficiently large. This just amounts to having a large chemical potential for the particles in the bound states at recombination.

\end{appendix}

\end{document}